%% file: main.tex
\documentclass[sigconf]{acmart}

\AtBeginDocument{%
  \providecommand\BibTeX{{%
    \normalfont B\kern-0.5em{\scshape i\kern-0.25em b}\kern-0.8em\TeX}}}

\setcopyright{none} % No copyright notice required for submissions
\renewcommand\footnotetextcopyrightpermission[1]{} % removes footnote with conference information in first column

\usepackage[T1]{fontenc}
\usepackage[utf8]{inputenc}
\usepackage{listings}
\usepackage{booktabs} % For formal tables
\usepackage{url}
\usepackage{mdframed}
\usepackage{pgfplots, pgfplotstable}
\usepackage{graphicx}
\usepackage{caption}
\usepackage{subcaption}
\usepackage{tikz}
\usetikzlibrary{calc}
\usetikzlibrary{patterns}
\usetikzlibrary{shapes.multipart, arrows}
\usetikzlibrary{decorations.pathreplacing}
\usetikzlibrary{shapes,snakes}
\usepackage{multirow}
\usepackage[PseudoCode]{algorithm}
\usepackage[noend]{algpseudocode}
\usepackage{color}
\usepackage{xcolor,colortbl}
\usepackage{enumitem}
\usepackage{balance}
\usepackage{filecontents}
\usepackage{amsthm}

\definecolor{cornellred}{rgb}{0.7, 0.11, 0.11}
\definecolor{arsenic}{rgb}{0.23, 0.27, 0.29}
\definecolor{auburn}{rgb}{0.43, 0.21, 0.1}
\definecolor{charcoal}{rgb}{0.21, 0.27, 0.31}
\definecolor{deepcarrotorange}{rgb}{0.91, 0.41, 0.17}
\definecolor{eggplant}{rgb}{0.38, 0.25, 0.32}

\definecolor{background}{HTML}{EEEEEE}
\definecolor{delim}{RGB}{20,105,176}
\colorlet{punct}{red!60!black}
\colorlet{numb}{red!60!black}

\def\mraft{\texttt{MRaft}}
\def\pbft{\texttt{PBFT}}
\def\rafttee{\texttt{Raft-TEE}}

\begin{document}\sloppy

\title{Mixed Fault Tolerance Protocols with \\Trusted Execution Environment}

\author{Mingyuan Gao}
\email{mingyuan.gao@u.nus.edu}
\orcid{0000-0002-9579-1890}
\affiliation{
  \institution{National University of Singapore}
  \country{Singapore}
}

\author{Hung Dang}
\email{hungdk4@fpt.com.vn}
\affiliation{
  \institution{FPT Blockchain Lab}
  \country{Vietnam}
}

\author{Ee-Chien Chang}
\email{changec@comp.nus.edu.sg}
\orcid{0000-0003-4613-0866}
\affiliation{
  \institution{National University of Singapore}
  \country{Singapore}
}

\author{Jialin Li}
\email{lijl@comp.nus.edu.sg}
\affiliation{
  \institution{National University of Singapore}
  \country{Singapore}
}

\begin{abstract}
\input{abstract}
\end{abstract}




\maketitle

%%% Submission-only Options
\pagestyle{plain} % removes running headers

\input{introduction}
\input{background}
\input{overview}
\input{mft}
\input{mraft}
\input{implementation}
\input{eval}
\input{related_work}
\input{conclusion}

\bibliographystyle{ACM-Reference-Format}
\bibliography{references}

\end{document}

%% file: abstract.tex
Blockchain systems, or distributed ledgers, are designed, built and operated 
in the presence of failures.
There are two dominant failure models, namely crash fault and Byzantine fault.
Byzantine fault tolerance (BFT) protocols offer stronger security guarantees, and thus are widely used in blockchain systems.
However, their security guarantees come at a dear cost to their performance and scalability.
Several works have improved BFT protocols, and Trusted Execution Environment (TEE) has been shown to be an effective solution.
However, existing such works typically assume that each participating node is equipped with TEE.
For blockchain systems wherein participants typically have different hardware configurations, 
i.e., \textit{some nodes feature TEE while others do not}, 
existing TEE-based BFT protocols are not applicable. 

This work studies the setting wherein not all participating nodes feature TEE, under which we propose a new fault model called \textit{mixed fault}, which is a combination of crash and Byzantine faults.
We explore a new approach to designing efficient distributed fault-tolerant protocols under the mixed fault model.
In general, mixed fault tolerance (MFT) protocols assume a network of $n$ nodes, among which up to $f = \frac{n-2}{3}$ can be subject to mixed faults.
Among these failures, some nodes may exhibit Byzantine behaviours, especially equivocating, while other nodes fail only by crashing.
We identify two key principles for designing efficient MFT protocols, namely, (i) prioritizing \textit{non-equivocating nodes in leading the protocol}, and (ii) advocating the use of public-key cryptographic primitives that allow \textit{authenticated messages to be aggregated}.
We showcase these design principles by prescribing an MFT protocol, namely \mraft, which is based on the Raft~\cite{raft} consensus protocol.

We implemented a prototype of \mraft~using Intel SGX, 
integrated it into the CCF \cite{ccf_tr} blockchain framework, 
and conducted experiments on Microsoft Azure using nodes spanning across different geographical regions.
Experimental results showed that MFT protocols can obtain the same security guarantees as their BFT counterparts while still providing better performance (both transaction throughput and latency) and scalability.

%% file: introduction.tex
\section{Introduction}
\label{sec:introduction}

% Distributed system and fault tolerance
Blockchain systems, or distributed ledgers, have received tremendous interest from both academia and industry communities over the last few years. 
They offer data integrity and immutability in the presence of service disruption or adversarial attempts. 
These systems achieve such security guarantees by building on distributed fault-tolerant consensus protocols~\cite{raft,pbft,sigmod_sharding}. 
These protocols can achieve both safety and liveness in the presence of failures. Safety means that honest participants (aka nodes or replicas) agree on the same value, whereas liveness means that these nodes eventually agree on a value. 

A fault-tolerant distributed system is designed, built and operated with respect to  a particular threat model, which comprises assumptions made on the nodes involved in upkeeping the system. 
Crash fault tolerance (CFT) protocols assume faulty nodes fail only by crashing, whereas Byzantine fault tolerance (BFT) protocols deal with faulty nodes that deviate arbitrarily from their protocol description. 
Byzantine faults cover not only adversarial behaviour, but also account for hostile environment wherein errors may arise from hardware malfunctions, software bugs, or errant system administrators causing data loss or state corruptions. 
There are clear trade-offs among choices of the threat model. 
BFT protocols offer stronger security guarantees in comparison to their CFT counterparts, for they are designed to tolerate  a more powerful adversary who is able to equivocate at will. It has been shown that equivocation - the act of a faulty node sending conflicting messages to other nodes without being detected - is the chief cause of the complications and overheads underlying BFT protocols~\cite{a2m}.  
Ensuring both safety and liveness despite equivocation comes at a dear cost to the performance (i.e., transaction throughput and latency) and scalability of BFT protocols \cite{blockbench}.

% Conventional and new approaches to fault tolerance
Consequently, a number of works have attempted to lessen this gap via the use of hybrid fault model~\cite{hypster, sigmod_sharding, a2m}. 
In such model, it is assumed that \textit{each node is equipped with a small trusted subsystem} that fails only by crashing, i.e., stop processing and responding to any message from other nodes, whereas other untrusted components in a node may fail or misbehave arbitrarily. 
Protocols following this hybrid fault model have been shown to require only $n = 2f+1$ nodes to tolerate $f$ Byzantine faults, as opposed to $n = 3f+1$ in conventional BFT protocols (e.g., PBFT~\cite{pbft}). This leads to lower computational and communication costs. 
Alternatively, there exist attempts that deploy CFT protocols such as Raft~\cite{raft} in a Byzantine setting~\cite{ccf_tr} via the use of Trusted Execution Environment (TEE)~\cite{enclave_formalization}. 
In particular, one may run the entire CFT consensus protocol inside a TEE with  attested execution~\cite{sgx} so as to curb adversarial capability of the faulty nodes. 
It is worth noting that this line of protocols assume that each node in the system is equipped with TEE, which may not be applicable to systems wherein participants have different hardware configurations, i.e., \textit{some nodes feature TEE while others do not}.

The motivations behind our work are twofold.
On one hand, participants of a fault-tolerant distributed system are likely to feature heterogenous machines, which are also subject to different maintainance or administration. Consequently, it is rarely the case that an adversary can coordinate a large number of Byzantine faults simultaneously as assumed in BFT threat models.
On the other hand, it is more likely the case that among the faulty nodes, \textit{only a few of them feature Byzantine behaviours, while others are simply crash faults}. 
The latter claim can be justified in practice via the use of TEE \cite{ccf_tr}, which is widely available nowadays in commodity processors. 
This is so because running code inside TEE can significantly restrict the malicious capability of a compromised node. 
That is, when deploying procotol code inside TEE, the TEE-powered node only fails by crashing. 

Given that BFT protocols are not efficient for such replicated systems, these two observations motivate us to design a new fault model, namely \textit{mixed fault}, which is a combination of crash and Byzantine faults.
In this fault model, a portion of these nodes may exhibit Byzantine behaviours, deviating arbitrarily from the protocol description in an attempt to break safety and liveness, whereas the remaining nodes fail only by crashing. 

% Our Model: Mixed Fault Tolerance - nodes with different trust assumption
In this work, we explore a new approach to designing efficient distributed fault-tolerant protocols under the mixed fault model, and we refer to such protocols as \textit{Mixed Fault Tolerance} (MFT) protocols.
In general, MFT protocols assume a network of $n$ nodes, among which up to $ f = \frac{n-2}{3} $ can be subject to mixed faults.
Among the $n$ nodes, some are equipped with TEE, while others do not.
% Principles that bring about efficiency: (i) reduce communication overhead, (ii) let the trustworthy do more
We identify two key principles for designing efficient MFT protocols. Both principles are drawn from an insight that communication overhead is a major hurdle that exacerbates protocols' scalability. 
Various works have demonstrated that communication overhead is the main performance bottleneck of these protocols~\cite{blockbench}. 

The first principle advocates \emph{prioritizing non-equivocating nodes in leading the protocol}. This is built upon an observation that the complications underlying BFT protocols primarily arise from equivocation.  
In MFT protocols, only a few faulty nodes exhibit equivocation, as opposed to each faulty node in BFT protocols. 
Consequently, as long as the protocol is led by a non-equivocating node, it may be designed in such a way that a majority of messages are routed through and verified by the leader, thereby reducing the overall communication and computation overheads in the system. 

The second principle advocates \emph{incorporating public-key cryptographic primitives that allow authenticated messages to be aggregated}. 
Early proposals for consensus protocols favour symmetric-key Message Authentication Code (MAC) to authenticate all-to-all communication in hostile environments, e.g., PBFT~\cite{pbft}. This choice was made so as to avoid public-key operations, whose implementations and executions used to be prohibitive. 
Fortunately, there exist various well-optimized asymmetric or hardware-based cryptosystems~\cite{gpu_asym_crypto, fast_signature, intel_manual} that render these costs far more affordable. 
Furthermore, by aggregating authenticated messages such that receiving and verifying an aggregated message is equivalent to receiving and verifying a set of individual messages, one can improve the communication complexity of the system~\cite{byzcoin, sigmod_sharding}.

To showcase these two principles, we prescribe an MFT protocol, namely, \mraft, which is based on a CFT consensus protocol called Raft~\cite{raft}.
It is worthy to mention that \mraft~ features communication complexity that is linear to the network size $n$, i.e., $O(n)$.
Similar to Raft, \mraft~is driven by a leader. However, the leader election in \mraft~favours nodes that are equipped with TEE. 
The TEE-powered leader leverages TEE to verify and aggregate messages from other nodes, generating a certificate attesting the fact that a statement has been agreed by a quorum of the nodes.
Alternatively, in the very rare case that the leader is not equipped with TEE, the protocol leverages Collective Signing (CoSi)~\cite{cosi} to generate such certificates.

We implemented a prototype of \mraft~using Intel SGX, 
integrated it into the CCF \cite{ccf_tr} blockchain framework, 
and conducted experiments on Microsoft Azure using nodes spanning across different geographical regions. 
Experimental results showed that MFT protocols can obtain the same security gurantees as their BFT counterparts while still providing better performance (both transcation throughput and latency) and scalability.

In summary, we make the following contributions in this work.
\begin{enumerate}
    \item Leveraging TEE, we propose a new approach to designing efficient distributed fault-tolerant protocols that tolerate a combination of crash and Byzantine faults. That is, a new fault model named \textit{mixed fault} is proposed.
    
    \item We identify two key principles for designing efficient MFT protocols. That is, (i) prioritizing non-equivocating nodes in leading the protocol, and (ii) advocating the use of public-key cryptographic primitives that allow authenticated messages to be aggregated.
    
    \item We showcase the above two design principles by prescribing an MFT protocol, namely, \mraft.
    
    \item We implemented a prototype of \mraft~ uisng Intel SGX, integrited it into the CCF blockchain framework \cite{ccf_tr}, conducted experiments in realistic deployment settings, and demontrated the efficiency of MFT protocols.
\end{enumerate}

%% file: background.tex
\section{Preliminaries}
\label{sec:prerequisite}

This section provides prerequisite knowledge that is relevant for this work.
We first discuss key features of distributed consensus protocols, focusing on Raft~\cite{raft} and PBFT~\cite{pbft}. 
Subsequently, we give a brief overview of TEE, in particular, Intel SGX.
Finally, we review the collective signing technique \cite{cosi}.

\subsection{Consensus Protocols}
\label{subsec:consensus}

Consensus protocols aim to achieve both \textit{safety} and \textit{liveness} in a distributed environment, which is potentially hostile. 
Safety necessitates non-faulty nodes to reach an agreement and never return conflicting results for the same query, whereas liveness requires that these nodes eventually agree on a value. 
There are two types of node failures, namely \textit{crash fault} and \textit{Byzantine fault}.  
Crash fault-tolerant (CFT) protocols assume faulty nodes fail only by crashing, whereas Byzantine fault-tolerant (BFT) protocols deal with faulty nodes that deviate arbitrarily from their  expected behaviours. For instance, a Byzantine node can equivocate, or delay its activity for arbitrary duration~\cite{pbft}.

\paragraph{Raft Consensus Protocol.}
Raft is arguably the most notable CFT consensus protocol. A Raft system comprises $n$ deterministic  nodes, and could tolerate up to  $f = \lfloor \frac{n-1}{2} \rfloor$ crash-failures. Each node maintains a log that contains a series of commands (or ledger). Raft ensures that logs of non-faulty nodes converge, achieving safety  regardless of synchrony assumption. However, it necessarily relies on timing to offer liveness~\cite{FLP_impossibility} (e.g., network is partially synchronous such that messages are delivered within an unknown but finite bound). 

The protocol is driven by a {\it leader}. All remaining nodes are referred to as  {\it followers}. 
Each leader is associated with a unique term. The leader exchanges heartbeats with the followers in order to maintain its leadership.
If a leader crashes, the protocol goes into the leader election phase and safely replaces the faulty leader with a non-faulty one. 
We refer readers to the Raft paper~\cite{raft} for details on the leader election. 

The leader collects commands (e.g., requests from the clients), records them in its log, and replicates them on the followers as follows. First, it broadcasts the command to all followers. Each command is identified by the leader's \textit{term} and an index in its log. When a follower receives a command from the leader, it appends the command  to its own log, and responds to the leader with an acknowledgement. The leader \textit{commits} (i.e., execute) the command  once it has received a quorum of  $f+1$  or more acknowledgements. The leader announces such commit to the followers, who then also commit the command in their own local state.

\paragraph{Practical Byzantine Fault Tolerance (PBFT)}
PBFT is driven by a leader, whose leadership is associated with a {\it view}. The protocol comprises three phases, namely $\texttt{Pre-Prepare}$, $\texttt{Prepare}$ and $\texttt{Commit}$. 
In the first phase, the leader collects requests from clients and broadcasts them to other nodes in the network as $\textsc{pre-prepare}$ messages. Upon receiving a $\textsc{pre-prepare}$ message from the leader, each node verifies the validity of the request, before broadcasting its responses in $\textsc{prepare}$ messages. These messages constitute the second phase, which ensures nodes agree on the ordering of the requests. 
Upon receiving a quorum of  valid and matching $\textsc{prepare}$ messages, nodes move to the third phase, broadcasting their  $\textsc{commit}$ messages. They execute the requests once they receive a quorum of $\textsc{commit}$ messages. When the leader fails, the view change  protocol is triggered to replace the leader.

PBFT requires a network of $n=3f+1$ nodes and a quorum size of $2f+1$ to tolerate up to $f$ Byzantine failures. The protocol observes a communication complexity of $O(n^2)$. It attains safety regardless of timing assumptions, whereas liveness is achieved in partially synchronous networks. 

\subsection{Trusted Execution Environment}
\label{subsec:SGX}

\paragraph{Enclave Execution.} 
Trusted Execution Environment (TEE) offers an isolated region that safeguards the integrity of the code running inside. In other words, an adversary is unable to tamper with the execution of the protected components, or deviate them from their expected behaviours. There are various  hardware primitives that provision TEEs, e.g., Intel SGX~\cite{sgx}, KeyStone~\cite{keystone} and Sanctum~\cite{sanctum}. This work adopts Intel SGX due to its wide availability.

Intel SGX~\cite{sgx} is capable of provisioning hardware-protected TEE (or \textit{enclave}) for general computation. 
Each enclave is associated with an address space (or enclave memory) which is guarded by the CPU, and inaccessible by
foreign (non-enclave) processes. In particular, each enclave is segregated from the Operating System (OS), user processes and other enclaves running on the same physical host. The enclave code, on the other hand, is able to invoke OS services such as paging and I/O. It is worth noting that data residing in the enclave memory are encrypted under the processor's key prior to leaving the enclave. 

\paragraph{Attestation.}
Enclaves are instantiated by the OS. A remote user can verify the correct instantiation of an enclave based on a remote attestation protocol~\cite{sgx_attestation_services}. The CPU produces a measurement of the enclave right after it is instantiated, and signs the measurement with its private key. Such measurement consists of  the hash of the enclave's initial state. The user can validate the signature using  Attestation Services~\cite{sgx_attestation_services}, and check the correctness of the measurement.

\paragraph{Data Sealing.} 
An enclave may persist its private state  on  non-volatile storage via data sealing mechanisms. Data sealing begins with the enclave obtaining a unique key bound to its measurement from the CPU. The enclave then encrypts its private state under the enclave-specific key before passing the encrypted data to the non-volatile storage. It is guaranteed that the sealed data is retrievable only  by its owner (i.e., the enclave that sealed it).
Nonetheless, previous works have shown that data sealing may be  susceptible to rollback attacks in which a malicious OS attempts to provide the enclave with properly sealed but stale data~\cite{rollback_detection}. Defences against such attack have been proposed in the literature~\cite{rote}.

\subsection{Collective Signing}
\label{subsec:CoSi}
Collective Signing, or CoSi for short, allows a group of independent nodes to validate and co-sign a statement~\cite{cosi}. The protocol produces a collective signature attesting the fact that all nodes in the group have endorsed the message. Such collective signature has size and verification cost equivalent to those of an individual signature. 

CoSi builds upon Schnorr multi-signatures~\cite{schnorr_sign}. The protocol takes advantage of communication trees~\cite{comm_tree1, comm_tree2} to optimize its communication cost, thereby achieving scalability. The protocol assumes that each node in the group has a unique public key, and that these keys are combined to generate an aggregate public key. One node in the group is designated as a leader, who drives the protocol through the following four phases to generate the collective signature for a message $M$:

\begin{itemize}
\item {\it Announcement:} The leader triggers the new round by multicast an announcement along the communication tree. 
$M$ may be embedded in the announcement. Alternatively, it can be sent in the {\it Challenge} phase. 

\item {\it Commitment:} Upon receiving the announcement from the leader, nodes pick a secret uniformly at random, and compute a Schnorr commitment of their chosen secret. From the bottom of the communication tree up, each node sends its aggregated Schnorr commitment to its parent. The node computes its aggregated commitment by combining its own Schnorr commitment with those it collects from its children. 

\item {\it Challenge:} After receiving the aggregated Schnorr commitment, the leader produces a collective Schnorr challenge. It then sends the challenge along the communication tree. If $M$ has not been sent in the {\it Announcement} phase, the leader embeds $M$ in the challenge. 

\item {\it Response:} Given the collective challenge, nodes assemble the aggregate responses in a manner similar to that of the {\it Commitment} phase.

\end{itemize} 

In case some nodes fail to respond to messages from the leader, the protocol can still produce the collective signature. However, this signature will include metadata that indicates which node did or did not participate in the collective signing. Readers are referred to the CoSi paper~\cite{cosi} for further details.

%% file: overview.tex
\section{Overview of MFT Systems}
\label{sec:overview}

In this section, we give an overview of the MFT systems.
First, we present some example distributed systems that motivate the design of the MFT model in Subsection~\ref{subsec:motivating_examples}.
Then, we describe the MFT model in detail in Subsection~\ref{subsec:system_model}.
Lastly, we elaborate on the threat model of an MFT system in Subsection~\ref{subsec:threat_model} and its system goals in Subsection~\ref{subsec:system_goals}.

\subsection{Motivating Examples}
\label{subsec:motivating_examples}

Before presenting the MFT model that this work studies, let's first look at some example distributed systems that motiviate the design of this model.

\begin{enumerate}[label=(E\arabic*)]
\setlength\itemsep{0.3em}

\item \textit{Consortium Blockchain:}
In a deployment of the consortium blockchain, the distributed ledger is shared and maintained by a group of independent parties. 
Typically, these parties rely on a BFT protocol like PBFT \cite{pbft} to provide safety and liveness for the distributed ledger. 
It is highly likely that these parties have different hardware configurations, e.g., some systems feature TEE while others do not.

\item \textit{Backward-Compatible Distributed Systems:}
Consider a large-sized corporate whose operations span across multiple regions, business activities at each region is administered by a separate branch. 
These branches need to stay in synchronization. 
Needless to say, the corporate can use a BFT protocol to enable such synchronization.
However, these branches are highly likely to have different hardware configurations. 
Some branches may have already upgraded their systems which feature TEE while other branches do not.

\item \textit{Confidentiality-Preserving Replicated Systems:}
Confidential Computing (CC) \cite{ccc} protects data in use by performing computation involving sensitive data in TEE, thus providing confidentiality protection for the sensitive data.
Since CC can increase the security assurances for sensitive and regulated data, various initiatives have been actively focusing on defining and accelerating its adoption.
When deployed in a replicated system, current CC platforms like CCF \cite{ccf_tr} typically assume each node in the system is equipped with TEE.
In circumstances where such assumption is too strong, i.e., not every node in the system feature TEE, 
one may still wish to attain the same security gurantee as the system configuration wherein all nodes feature TEE.
In such cases, the goal can be achieved by using the network only for reaching consensus on the order of execution, while the actual execution of the confidential computation is carried out on the TEE-powered nodes. The computation results are then replicated to other nodes.

\end{enumerate}

The above distributed systems all aim to provision \textit{a replicated service using a BFT protocol}.
One one hand, \textit{nodes in these systems are likely to feature heterogenous machines}, which are also subject to different maintainance or administration. Consequently, it is rarely the case that an adversary can coordinate a large number of Byzantine faults simultaneously as assumed in BFT threat models.
On the other hand, it is more likely the case that among the faulty machines, \textit{only a few of them feature Byzantine behaviours, while others are simply crash-faults}. 
This claim can be justified in practice via the use of TEE \cite{ccf_tr}, which is widely available in recent commodity processors. 
This is so because running code inside a TEE can significantly restrict the malicious capability of a compromised node.
That is, \textit{when deploying procotol code inside a TEE, the TEE-powered node only fails by crashing.}

Given that BFT protocols are not efficient for such replicated systems, the above two observations motivate us to design a new fault model for such distributed systems.

\subsection{System Model}
\label{subsec:system_model}

We now present the system model that this work studies, with a focus on the fault model.

We study a distributed system that comprises of $n$ deterministic and independent nodes.
The system provisions a replicated service that receives requests from individual clients, and executes those requests in a totally ordered sequence. In other words, the replicated service appears to the clients as if it runs on a single non-faulty machine. 

Most distributed systems make an assumption that nodes in the network are homogeneous.
That is, the nodes are presumed to share the same set of capabilities, and admit similar vulnerabilities. 
Nonetheless, this is not always necessarily the case. 
As can be seen from the above example systems, nodes are likely to have heterogeneous machines.
\textit{In view of this and the wide availability of TEE in commodity processors, we study a heterogeneous system wherein some nodes feature hardware-based TEE}, 
while others are powered by legacy systems which place trust on their OSs or hypervisors.

Each pair of nodes in the network communicate through a reliable, authenticated point-to-point communication channel. 
In order to sidestep the FLP impossibility~\cite{FLP_impossibility}, we assume that the communication channels are partially synchronous, i.e., messages that are sent repeatedly with a finite timeout will be eventually delivered at its destination. 
This assumption is commonly observed in existing distributed and replicated systems~\cite{sigmod_sharding, byzcoin}. 
Besides, there is no global clock. Nodes process messages and execute requests at their own speed.

\paragraph{Mixed Fault Model:} 
It has been shown that the complexity of BFT protocols typically arises from the ability of a Byzantine node to equivocate (i.e., issue conflicting statements to different nodes without being detected) \cite{a2m, sigmod_sharding}.
Given that code executed inside a TEE is integrity protected, the malicious behaviours of a compromised node that is powered by TEE can be significantly restricted. 
That is, when deploying protocol code inside the TEE, a TEE-powered node, even compromised, is not able to equivocate. 
In other words, a TEE-powered node only exibits crash failures.
Thus, the use of TEE can significantly reduce the communication complexity of BFT protocols.

In heterogeneous distributed systems wherein some nodes feature TEE while others do not, different nodes can exibit different types of failures. In particular, TEE-powered nodes \textit{only} exhibit \textit{crash} failures, while non-TEE nodes feature \textit{Byzantine} behaviours. We refer to this fault model as the \textit{mixed fault}, which is eleborated in Section~\ref{sec:mft}.

\subsection{Threat Model}
\label{subsec:threat_model}

Our threat model assumes that the Byzantine nodes are under adversarial control. They may access (i.e., read and write) to other processes' memory, including that of the OS. They can also tamper with data persisted on persistent storage, intercept and alter system calls. 

The adversary is also able to initialize, stop and invoke the TEE enclaves of the TEE-powered nodes. Nevertheless, its control over the TEE is limited, for we make an assumption that the TEE's attested execution mechanism is secure. 
In contrast to Intel SGX's threat model, we make no assumption on the confidentiality protection of the enclaves, except for a few critical cryptographic primitives such as key generation, random number generation or attestation. 
That is, the TEE-powered nodes run in the seal-glassed proof model that is able to attest to the correct execution of the codebase loaded inside, but its execution is transparent~\cite{sealedglass}. This threat model is particularly relevant in view of recent side-channel attacks against Intel SGX (e.g., \cite{sgx_cache_attack}). While we leave attacks that compromise confidentiality of attestation and other cryptographic keys~\cite{foreshadow} out of scope, we remark that techniques, both software and hardware-based, hardening critical cryptographic operations against these attacks are available~\cite{kiriansky2018dawg}. 

Finally, we assume the adversary is computationally bounded. It is not able to break standard cryptographic assumptions. Besides, we exclude denial-of-service attacks against the system in this study.

\subsection{System Goals}
\label{subsec:system_goals}

Our system goal is efficient state machine replication \cite{state_machine_approach} under the MFT model.
It is desired that the system provides both \textit{safety} and \textit{liveness}. That is, any two clients interact with the system receive consistent responses, and valid requests from the clients are eventually executed. 

In particular, we study a network of $n$ nodes, among which up to $f = \frac{n-2}{3} $ can be subject to mixed faults. 
The argument for this is provided in Section~\ref{subsec:mraft_leader_election}.
Here, $f=f_c+f_b$, wherein $f_c$ denotes the number of crash-faulty nodes, and $f_b$ that of Byzantine nodes.
In addition, we require that the number of TEE-powered nodes $n_{tee} \geq f+1 $.
We refer to protocols that enable such replication as \textit{MFT protocols}, which are elaborated in Section~\ref{sec:mft}.

%% file: mft.tex
\section{Mixed Fault Tolerance}
\label{sec:mft}

In this section, we first elaborate on our MFT model by contrasting it against related conventional and non-conventional fault tolerance models, thereby highlighting MFT's key characteristics. For clarity, we shall review those models as we visit them. 
Then, we present two design principles that allow MFT protocols to scale.

\subsection{MFT vs. CFT/BFT}
Recall that our system model presented in Section~\ref{subsec:system_model} does not assume any global clock or known bounds of network latency. This assumption is also observed by asynchronous CFT/BFT protocols. The difference between MFT and asynchronous CFT/BFT arises from our treatment of node faults.

\input{figures/MFT_vs_hybrid_faults}

CFT protocols, such as Raft~\cite{raft}, provide safety regardless of network condition, and require partial synchrony to ensure liveness. They tolerate up to $f = \frac{n-1}{2} $ crash failures. However, as soon as there exists a single Byzantine node in the network, CFT's threat model is violated, and all security guarantees are voided.
In contrast to CFT, \textit{our MFT model affords some Byzantine nodes, and retains safety and liveness as long as the number of faulty nodes does not exceed a predefined threshold}.

BFT protocols assume a powerful adversary who wields absolute control over all faulty nodes, causing them to deviate arbitrarily from their expected behaviours. 
The most prominent BFT  protocol is arguable PBFT~\cite{pbft}, which tolerates up to $f = \frac{n-1}{3} $ faults but incurs quadratic communication complexity in term of the network size. Such communication complexity hinders PBFT's scalability~\cite{sigmod_sharding, blockbench}. 
In opposition to BFT, \textit{MFT assumes that only a portion of the faulty nodes exhibit Byzantine behaviours}, whereas other faulty nodes only crash and do not misbehave.
\textit{This assumption allows MFT to trim down the communication overhead}, thereby improving the performance and scalability of the system.

\subsection{MFT vs. Hybrid BFT}

\textit{The complexity of BFT protocols typically arises from the ability of a Byzantine node to equivocate} 
\cite{a2m, sigmod_sharding}. It has been shown that without equivocation, it is possible to tolerate $f$ Byzantine failures with only $n = 2f + 1$ nodes using the quorum size $f + 1$. 
The smaller network and quorum sizes result in lower computational and communicational cost incurred in tolerating the same number of failures.

Building on this observation, a number of approaches have studied the hybrid fault model.
In such models, each node in the network is assumed to be equipped with a trusted subsystem that only fails by crashing, whereas its other components are untrusted and may fail arbitrarily. 
Figure~\ref{subfig:hybrid} depicts this threat model. The trusted subsystem is utilized to combat against undetected equivocation. A common technique is to bind each message a node broadcasts with a record in a log (which can be as simple as a monotonic counter) maintained by the trusted subsystem. Since operations carried out by the trusted subsystem cannot be equivocated, malicious nodes cannot send conflicting messages without being convicted by others.

Alternatively, one can also eliminate equivocation by running the entire codebase of a consensus protocol inside a TEE (Figure~\ref{subfig:TEEconsensus}). 
This approach, adopted by CCF~\cite{ccf_tr}, effectively reduces a node's fault model from BFT to CFT. Consequently, any non-Byzantine consensus protocols, such as Raft~\cite{raft} or Paxos~\cite{paxos}, can be applied. 
CFT consensus protocols could then tolerate $f$ Byzantine failures with only $n = 2f + 1$ nodes using the quorum size of $f + 1$, resulting in lower communication and computational overhead.
While this approach is similar to the hybrid fault model described earlier in their assumptions on the availability of TEE at each node, it incurs a large trusted code base (TCB) which is undesirable for security~\cite{minimal_TCB}.
A large TCB makes security analysis of the implementation bewildering, which likely exposes the system to potential vulnerabilities.

In contrast to the above two hybrid fault models, \textit{MFT does not require each node in the network to be equipped with TEE}.
Our MFT model allows for a portion of the nodes to behave arbitrarily (i.e., Byzantine nodes), whereas hybrid fault models collapse as soon as there exists a single Byzantine node in the network. Figure~\ref{subfic:MFTreplica} shows the heterogeneity of nodes in an MFT system.

\subsection{MFT vs. Flexible BFT}
Malkhi et al.~\cite{flexiblebft} introduced Flexible BFT, which tolerates alive-but-corrupt faults. Flexible BFT assumes that these alive-but-corrupt nodes may exhibit Byzantine behaviours.
However, they do so strictly for the purpose of breaking the protocol's safety. In case they are unable to compromise safety, they will not hinder the protocol's liveness. Furthermore, Flexible BFT allows clients interacting with the replicated service to hold different assumptions or beliefs about the system, based on which they interpret the protocol transcript and make commit decisions. Flexible BFT guarantees both safety and liveness for all clients with correct beliefs.

Flexible BFT justifies the alive-but-corrupt faults based on an observation that the adversary may benefit if safety is broken, for instance in double-spending attacks, while it is unlikely to gain anything from broken liveness. 
The authors~\cite{flexiblebft} further argue that alive-but-corrupt nodes are incentivised to keep the liveness as they could collect service fee. These assumptions are in line with rational protocol design treatment wherein the adversary is assumed to misbehave only if such action yields (economic) gain 
~\cite{rational_adv}. 

On the contrary, MFT does not make any assumption on the rationale of the corrupted nodes. Our threat model pays more attention to the capability that the adversary wields and the constraints that it adheres to. For instance, if a TEE's attested execution protection is intact, the adversary may attempt to disconnect it from the network, but it cannot compromise TEE's execution integrity. In such a case, the adversary cannot cause the TEE-powered nodes to violate safety, yet it can tamper with their I/O and network connections in an attempt to prevent liveness.

\subsection{MFT vs. XFT}
Cross Fault Tolerance, or XFT for short ~\cite{xft}, studies a system model which admits both crash and Byzantine faults. Beyond crash and Byzantine faults, XFT explicitly defines network fault as an event wherein some  non-faulty nodes could not communicate synchronously with each other (i.e., a message exchanged between two nodes is delivered and processed within a known latency $\Delta$). A node is considered partitioned if it does not belong to a largest synchronous subset. 
In a network of $n$ nodes, XFT protocols are able to tolerate up to $n$ crash faults without compromising safety, and tolerate some Byzantine faults together with network faults, so long as there exist a majority of nodes that are not faulty  and communicate synchronously.

The key difference between XFT and our MFT model is XFT's separation of node and network faults. 
Similar to CFT and BFT, our MFT model considers only machine faults, and relies on partial synchrony to sidestep the FLP impossibility~\cite{FLP_impossibility}.
By separating network from node faults, XFT can guarantees safety in two modes:
(i) there is no Byzantine faults, regardless of the number of crash-faulty and partitioned replicas;
(ii) there exist some Byzantine faults, but a majority of nodes remain correct and not partitioned, 
i.e., the total number of crash, Byzantine and network faults combined does not exceed $ \frac{n-1}{2}$.
MFT, on the other hand, offers safety when $f \leq  \frac{n-2}{3}$ ($f$ is the total number of crash and Byzantine nodes) in partially synchronous network.

\subsection{Design Principles for MFT Protocols}
So far we have contrasted MFT against CFT, BFT, and other non-conventional fault tolerance models. We now draw  observations from the discussion and comparison presented above, and codify them into two design principles. 

Our first observation, which is applicable not only to MFT but also to other distributed systems in general, is that communication complexity is typically a bottleneck to the performance and efficiency of the system. Various works have shown that this is indeed the case for many different fault tolerance models and protocol designs~\cite{blockbench}. Consequently, reducing the communication complexity contributes to the improvement of the system's performance.

MFT protocols essentially implement a replicated state machine system that is driven by a designated node in the network, which is typically called \textit{leader}. That is, the leader processes requests sequentially and the remaining nodes in the network merely follow. When the leader becomes faulty, it is replaced by another node via a subprotocol, during which the performance of the system suffers significantly~\cite{sigmod_sharding}. 
Our second observation is that in heterogeneous settings such as those studied under the MFT model, the choice of the leader plays a crucial role in tuning the performance of the system.

These two observations motivate the following two design principles for MFT protocols:

\begin{enumerate}[label=(P\arabic*)]
\setlength\itemsep{0.3em}

\item {\it Leadership favours non-equivocating nodes.}
Recall that MFT assumes both types of failures, namely crash and Byzantine faults. It is well established that most complications underlying BFT protocols are due to equivocation.
In MFT, only Byzantine nodes may equivocate, whereas such behaviour is never conducted by crash faulty nodes.  
Consequently, as long as a protocol is led by a non-equivocating node, a majority of consensus messages can be routed through and verified by the leader on behalf of other nodes.  
This communication pattern poses much less overhead on the network. Furthermore, computation cost incurred in verifying authenticated messages can also be saved. 
We remark that this principle advocates giving higher priority to non-equivocating node in attaining the leadership, but it does not impose strict restriction on potentially Byzantine node never becoming the leader.

\item {\it Aggragation of Consensus Messages.}
In case the leadership is taken by a node that may feature equivocation, it is important to ensure that its misbehaviour, if any, does not compromise safety. Such node should not be trusted with collecting and verifying consensus messages on others' behalf. Instead, the protocol should incorporate cryptographic primitives that allow consensus messages to be efficiently aggregated in such a way that receiving and verifying an aggregated message is equivalent to receiving and verifying a quorum of individual messages (e.g., CoSi~\cite{cosi}), without relying on any trusted third party. 
This clearly improves the communication complexity of the system. 

\end{enumerate}

Based on these two principals, we retrofitted the Raft procotol for the MFT model, and refer to the resulting protocol as \mraft, which is elaborated in Section~\ref{sec:mraft}.

%% file: figures/MFT_vs_hybrid_faults.tex
\begin{figure*}[t!]
\begin{subfigure}{0.33\textwidth}
\centering
{\includegraphics[width=0.92\textwidth]{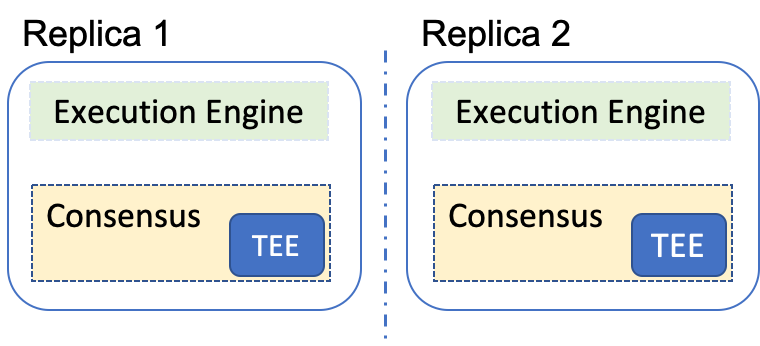}}
\caption{Hybrid Fault Replicas.}
\label{subfig:hybrid}
\end{subfigure}
\hfill
\begin{subfigure}{0.33\textwidth}
\centering
{\includegraphics[width=0.92\textwidth]{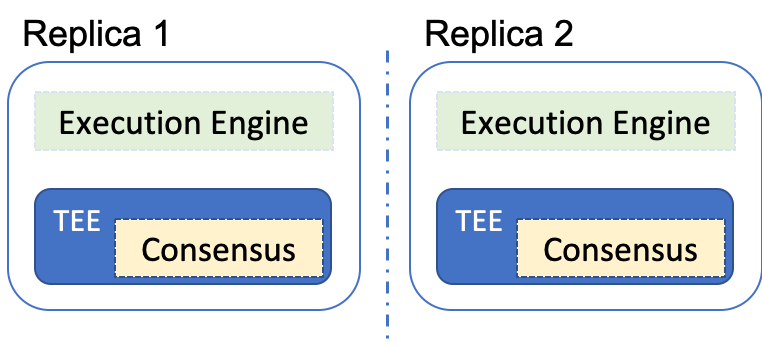}}
\caption{TEE-Powered Consensus.}
\label{subfig:TEEconsensus}
\end{subfigure}
\hfill
\begin{subfigure}{0.33\textwidth}
\centering
{\includegraphics[width=0.92\textwidth]{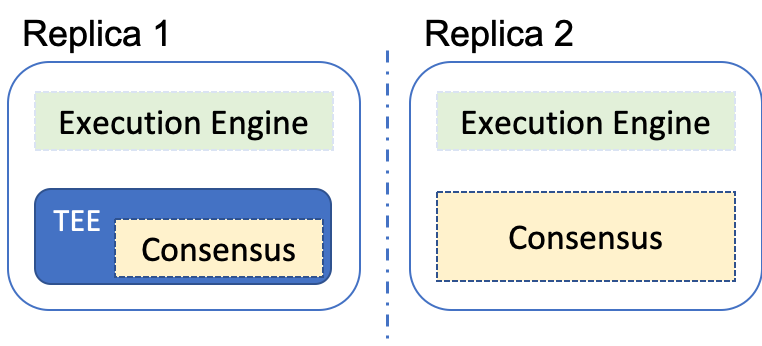}}
\caption{MFT Replicas.}
\label{subfic:MFTreplica}
\end{subfigure}
\caption{Comparison between MFT and other hybrid fault models.}
\label{fig:MFT_vs_hybrid}
\end{figure*}

%% file: mraft.tex
\section{\mraft}
\label{sec:mraft}

In this section, we present \mraft, which is a distributed consensus protocol designed to operate under the MFT model. 
\mraft~ is built upon Raft \cite{raft}. 
However, instead of tolerating up to $f$ crash-faults using a network of $n=2f+1$ nodes as Raft does, \mraft~employs a network of $n=3f+2$ nodes to tolerate up to $f$ mixed faults. 
Some nodes in the network are assumed to be equipped with TEE with intact integrity protection, and thus never equivocate or deviate from the protocol description. 

We require that the number of TEE-powered nodes $n_{tee} \geq f+1 $. This assumption is not strictly required, but it significantly increases the chance that a TEE-powered node is elelected as the leader. We further assume that each node should know TEE-capability of other nodes in the network, i.e., whether the other node is equipped with and running the consensus codebase inside a TEE, which can be achieved via TEE's remote attestation mechanism~\cite{sgx_remote_attest}.

Each node in the network implements a state machine, and maintains a replicated \textit{log} which records a sequence of commands or requests the network has served. 
The goal of \mraft~ is to ensure logs of different nodes in the network converge, and  each node commits (or executes) exactly the same sequence of commands.

Similar to Raft, nodes in \mraft~can be in one of the three roles, namely \textit{leader}, \textit{follower} and \textit{candidate}. Our protocol proceeds in terms. In each term, a node is selected to serve as a leader, while all other nodes are followers. 
Leader election in \mraft~ favours TEE-powered nodes, which are assumed to be non-equivocating thanks to TEE's execution integrity protection. In the common case where a TEE-powered node attains leadership, the message pattern and communication complexity of \mraft~is similar to those of Raft. The leader processes most of the messages, while the followers passively receive and respond to messages from the leader.

In a very rare case where all TEE-powered nodes fail to attain leadership, the protocol is led by a non-TEE node. In such case, the leader cannot be trusted to verify followers' messages on behalf of the network.
The first approach to sidestep this issue is to employ an all-to-all communication pattern wherein a node broadcasts its messages to the network (similar to the message pattern of PBFT~\cite{pbft}). 
However, this will results in a communication complexity of $O(n^2)$, which is detrimental to the scalability of the protocol~\cite{sigmod_sharding}. 
Alternatively, one can tasks the leader to collect messages containing digital signatures from all followers. Once it has collected a quorum of signatures, it broadcasts such quorum to the network, incurring a communication complexity of $O(n)$ instead of $O(n^2)$. 
Nonetheless, the need of each node to independently verify quorum of signatures may poses a hindrance on the performance of the protocol. 
In view of these encumbrances, we follow Byzcoin~\cite{byzcoin} in using CoSi~\cite{cosi} to implement collective signing. This implementation enables a potentially equivocating leader to collect and aggregate messages from the followers. It suffices for a follower to receive and verify an aggregated message before proceeding, as opposed to verify a quorum of messages as in the two naive approaches mentioned earlier.

We detail below the concrete prescription of \mraft~in a common case wherein the protocol is led by a TEE-powered node (Section~\ref{subsec:mraft_tee_leader}) and in a very rare case wherein the non-TEE node attains leadership (Section~\ref{subsec:mraft_non_tee_leader}). For clarity of exposition, we shall denote the leader by $L$, and a follower $i$ by $p_i$. 
Each leader $L$ is associated with a term $t$, and indexes a request $r$  it receives from the client in its log with a counter $j$. Both the term number $t$ and the index $j$ are increased monotonically.  
That is, each request $r$ is uniquely identified by a 3-tuple $\langle t, j, r \rangle$.
In addition to the replicated log, each node keeps track of a $\texttt{LastCommitIndex}$ which is the  index of the latest entry in its own replicated log that is known to be committed, which also increases monotonically over time.
With each $p_i$, $L$ establishes an authenticated communication channel, and a timeout $T_i$ which is a window of time during which $p_i$ expects to receive a message from $L$.  Should there be no request from the client during the timeout window $T_i$, $L$ sends a $\texttt{HeartBeat}$ message containing $L$'s term number and $\texttt{LastCommitIndex}$ to $p_i$ so as to maintain its leadership. 
If $p_i$ fails to hear from $L$ once its timeout $T_i$  has passed, it assumes $L$ is faulty and requests for a new leader. We defer the leader election mechanism to Section~\ref{subsec:mraft_leader_election}.

\subsection{TEE-Powered Leader}
\label{subsec:mraft_tee_leader}
\paragraph{Protocol.}
If the leader $L$ is equipped with TEE and running the protocol inside the TEE, it is assumed that $L$ never deviates from its expected behaviour (i.e., execution integrity is preserved). It may fail only by crashing. In such case, the protocol proceeds as follows:

\begin{enumerate}
\item Upon receiving a request $r$ from a client, the leader $L$ whose term is $t$ assigns $r$ an index $j$, and puts $\langle t, j, r \rangle$ onto its log.

\item $L$ broadcasts $\langle t, j, r \rangle$ to the network.

\item Upon receiving $\langle t, j, r \rangle$, a follower $p_i$ checks if it was from the node it believes to be the leader, and  if $r$ is valid (i.e., committing $r$ does not compromise safety). If so, it puts $\langle t, j, r \rangle$ to its log and responds $L$ with an acknowledgement $\langle \texttt{ack}_i, t, j, r \rangle$.

\item Upon receiving a quorum of $q=2f+1$ acknowledgements for $r$, the leader commits $r$ (i.e., executes $r$ and applies the result to its state machine) and all uncommitted requests in its log whose index is smaller than that of $r$, if any. 
It produces a certificate $\texttt{Cert}_r$, which attests a fact that $r$ has been replicated on a quorum of nodes. Subsequently, $L$ broadcasts  $\texttt{Cert}_r$ to the network.

\item Upon receiving $\texttt{Cert}_r$ from $L$, a follower $p_i$ commits $r$ and all uncommitted requests in its log whose index is smaller than that of $r$, if any.

\end{enumerate}

\paragraph{Remarks.}
Our protocol requires the leader $L$ to broadcast $\texttt{Cert}_r$ in its announcement of request $r$'s commit (Step 4). This enables every node in the network to independently verify that $r$ has been replicated on a quorum of nodes, and that the quorum agrees on the total order of requests. 
Since quorum size in our protocol is  $q=2f+1$, quorums must be intersected at at least one honest node. 
Consequently, while a Byzantine node could equivocate, or tamper with term and index value in its messages, 
it is unable to cause the entire network to violate safety. 

\subsection{Leader Election}
\label{subsec:mraft_leader_election}
As mentioned earlier, during normal operation, $L$ periodically sends $\texttt{HeartBeat}$ 
message containing its term number and $\texttt{LastCommitIndex}$ to followers so as to maintain its leadership.
Communication during the consensus round can also be deemed as $\texttt{HeartBeat}$ 
messages, for it conveys $L$'s term number and $\texttt{LastCommitIndex}$, as well as the fact that $L$ is fully functional.
When a follower $p_i$ receives such messages, it  acknowledges with a corresponding
$\texttt{ack}$ via an authenticated channel. 
$p_i$ may rely on $L$'s $\texttt{LastCommitIndex}$ to assure that its
replicated log and state are in sync with $L$'s.
Should $p_i$ finds its log and state outdated,  it retrieves the missing requests (i.e., log entries) 
and commits them, thereby updating its own state to match that of $L$.

For each follower $p_i$, $L$ institutes a unique and randomised timeout $T_i$ chosen from a fixed interval. 
Since \mraft~favors TEE-powered node in obtaining leadership, the timeout interval (i.e., a fixed interval from which $T_i$ is drawn) between $L$ and a TEE-powered node, say $[T_a, T_b]$ is configured to be smaller than that between $L$ and a non-TEE follower, say $[T_c, T_d]$. That is, $T_a < T_c \wedge T_b < T_d$.

Should a follower $p_i$ fail to receive any message from $L$ after its $T_i$ has elapsed, it switches its role to candidate and increases its term number. Subsequently, $p_i$ broadcasts a $\texttt{RequestVote} = \langle p_i, newTerm, lastLogIndex \rangle$ to the network in an attempt to assume leadership, in which $newTerm$ is its current term number, and $lastLogIndex$ is the index of the latest entry in its replicated log. A leader election in  \mraft~is closely related to that of Raft, with some adjustments in how a recipient handles a $\texttt{RequestVote}$ message. 

Upon receiving a $\texttt{RequestVote}$ from $p_i$, a node $p_j$ grants its vote if all following conditions are met:

\begin{itemize}
\item $p_i$ is indeed a node within the network, and $\texttt{RequestVote}$ is properly signed by $p_i$
\item $p_j$'s own timeout $T_j$ has elapsed and it has not received any message from its current leader.
\item $newTerm$ in the $\texttt{RequestVote}$ is larger than its own current term.
\item $p_j$ has not granted its vote to any other candidate.
\item $p_i$'s replicated log is more up-to-date than that of $p_j$, as determined by the indice of the last entries.
\end{itemize}

In case $p_j$ receives a $\texttt{RequestVote}$ from $p_i$ before its timeout $T_j$ with its current leader has not elapsed, 
it queues $p_i$'s $\texttt{RequestVote}$, if it has not queued any other candidate's $\texttt{RequestVote}$, and $p_i$'s log 
is more up-to-date than its own replicated log.
When there is a competing $\texttt{RequestVote}$, $p_j$ keeps that of a node which is more up-to-date, and discards the 
other. 
If $p_j$ receives a valid message from its current leader, it discards any $\texttt{RequestVote}$ that it has queued.
When $T_j$ has elapsed, $p_j$ grants its vote to $p_i$. 

A vote is essentially an authenticated message from $p_j$ that is publicly verifiable. A node wins an election once it has 
collected votes from a quorum of $q=2f+2$ nodes. If the leader is equipped with TEE, it can leverage the TEE to produce a 
compact proof of leadership by aggregating the votes, as in step (4) of the protocol described in Section~
\ref{subsec:mraft_tee_leader}. On the other hand, if a non-TEE candidate wins the election, the proof of its leadership is a 
collection of the votes it has thus received. The new leader announces its authority by broadcasting a $\texttt{HeartBeat}$  
message containing its proof of leadership along with the new term  number to the network. 
Upon receiving such message, a follower verifies if the proof of leadership is valid before switching to the new leader and 
updating its term number accordingly.

\paragraph{Remarks.}
We remark that the quorum size necessitated for leader election is $q=2f+2$. Hence, the network needs $n=3f+2$ nodes to be correctly operational.
This is so because there exists a possibility that an honest node is disconnected from the network during the leader election. In such a scenario, it may happen that there are $f$ Byzantine followers, $f$ honest and up-to-date followers,  $f$ honest followers that hold stale view of the replicated log, and a candidate that may have a stale view. If the quorum size for leader election is $q=2f+1$, it may happen that the log of elected leader miss some entries committed by the previous leader, and his leadership in the new term may accidentally undo the requests that had been committed earlier. 
To avoid this, the candidate must obtain $q=2f+2$ votes. Therefore, the network needs $n = 3f+2$ nodes to tolerate $f$ failures.

\subsection{Non-TEE Leader}
\label{subsec:mraft_non_tee_leader}

Note that \mraft~employs a network of $n=3f+2$ to tolerate up to $f$ mixed faults, among which up to $f$ can be Byzantine faults.
Since we require that the number of TEE-powered nodes $n_{tee} \geq f+1 $, thus, it is high likely that a TEE node becomes leader during the leader election process.
In a very rare case that the leader election fails to elect a TEE-powered node, we can repeat the leader election process until a TEE-powered node is elected as the leader. 
However, this approach would sacrifice the system's liveness during the leader election process. 
Alternatively, we can set a timeout value for the period that the system is electing a TEE-powered node as leader.
When this timeout is reached, we can temporarily fall back to a safe BFT protocol until a TEE node is available and elected as leader.
That is, we use the fallback BFT protocol to ensure that the system's \textit{liveness} is lost only during the timeout period.

Below is the fallback protocol we prescribed for the circumstance when a TEE-powered node is temporarily not eleteced as the leader.

\paragraph{Protocol}
When the leader is not equipped with TEE, it cannot be trusted to aggregate responses from the followers as in the previous case. Consequently, \mraft~employs CoSi to save on the communication complexity. The protocol proceeds as follows:

\begin{enumerate}
\item Upon receiving a request $r$ from a client, the leader $L$ whose term is $t$ assigns $r$ an index $j$, and puts $\langle t, j, r \rangle$ onto its log.

\item $L$ initiates a CoSi round to drive the network to generate the collective signature for a message $\langle t, j, r \rangle$. A successful CoSi round effectively replicates $\langle t, j, r \rangle$ on the followers' logs.

\item Upon receiving $\langle t, j, r \rangle$, during the execution of the CoSi protocol (described in Section~\ref{subsec:CoSi}), a follower $p_i$ checks if it was from the node it believes to be the leader, if $r$ is valid, and if the term $t$ and index $j$ match its log ($j$ should immediately follow  the latest committed entry in its log). If so, it puts $\langle t, j, r \rangle$ to its log and completes the final phase of the CoSi protocol.

\item Once the network has completed the CoSi rounds, $L$ should have obtained the collective signature $\texttt{CoSig}_r$ for $\langle t, j, r \rangle$. It checks if a quorum of $q=2f+1$ has partaken in the collective signing using the metadata 
contained in the collective signature. If this is the case, it is assured that the $\langle t, j, r \rangle$ has been replicated on a 
quorum of nodes, and it is safe for $L$ to commit $r$ and all uncommitted requests in its log whose index is smaller than that of $r$, if any. Subsequently, $L$ broadcasts  $\texttt{CoSig}_r$ to the network.

\item Upon receiving $\texttt{CoSig}_r$, a follower checks if a quorum of $q=2f+1$ nodes have co-signed $\texttt{CoSig}_r$. If so,  $p_i$ commits $r$ and all uncommitted requests in its log whose index is smaller than that of $r$, if any.

\end{enumerate}

\paragraph{Remarks}
In case the leader $L$ is equipped with a TEE, \mraft~relies on the trusted execution to produce a certificate $\texttt{Cert}_r$ which attests a fact that a request $r$ has been replicated on a quorum of nodes. In case $L$ does not feature TEE, \mraft~resorts to the CoSi protocol (and relies on its security) to produce $\texttt{CoSig}_r$, which conveys the same significance that $\texttt{Cert}_r$ does. 

%% file: implementation.tex
\section{Implementation}
\label{sec:implementation}

In this section, we describe the implementation details of our prototype~\mraft. Our prototype is based on the codebase of CCF \cite{ccf_tr}, which is an open-source framework for building confidential replicated services.

A CCF network \cite{ccf_doc} consists of several nodes, each running on top of a TEE, such as Intel SGX. 
Each node runs the same application, 
which can mutate or read the in-enclave-memory \emph{key-value store} that is replicated across all nodes in the network. 
The key-value store is a collection of maps (associating a key to a value) that are defined by the application. 
Changes to the key-value store must be \emph{agreed} by a \emph{quorum} number of nodes before being applied, wherein the quorum value depends on the \emph{consensus} algorithm selected. 

CCF supports two consensus protocols, i.e., CFT and BFT. CFT is the default consensus protocol and its implementation is based on Raft \cite{raft}.
The BFT implementation is a derivative from PBFT-PK (PBFT using signatures) \cite{pbft}, with additional features specific to CCF \cite{shamis2021pac}. 

Each CCF network has a network identity public-key certificate (aka, \emph{service certificate}), used for TLS server authentication, and the corresponding \emph{private key} always resides in enclave memory. This key pair is generated when the first node starts.
Each CCF node is identified by a fresh \emph{public-key certificate} endorsed by the enclave quote. This \emph{node-identity certificate} is used to authenticate the node when it joins the network, and to sign entries committed by the node to the ledger during its time as primary.

\paragraph{Modifications to CCF's Codebase} 
A CCF network only allows TEE nodes with a valid enclave quote to join the network, which is achieved by verifying a joining node’s enclave quote through remote attestation. 
For non-TEE nodes, we similary identify them using a public-key certificate issued by the \emph{service certificate}, i.e., the service certificate acts as the root CA for these node-identity certificates.
We retrofitted CCF's codebase to allow non-TEE nodes with a valid \emph{node-identify certificate} to join the network.

In CCF, each node to node pair establish a symmetric traffic key, using an authenticated Diffie-Hellman key exchange. This key is used to authenticate messages sent between nodes. 
For messages sent from non-TEE nodes, we retrofitted CCF's codebase to append a signature to such messages; the signatures are generated using the private key corresponding to the \emph{node-identity public key}. Thus, TEE nodes can verify the authenticity of such messages using the contained signature.

\paragraph{Changes to CCF's Raft Implementation}
To implement \mraft, we retrofitted the Raft implementation in \emph{ccf-1.0.0}, and 
changed its \emph{quorum} size to $2f+2$, wherein $f=\frac{n-2}{3}$.
Specifically, we modified the code so that the TEE-powered leader institutes smaller leader-election timeout values for TEE nodes than for non-TEE nodes, which makes leader election favors TEE nodes. 
We also modified the code so that, during normal operation, whenever a TEE-powered leader commits entries, 
it generates a commit certificate for these entries. 
Followers will update their \texttt{commit\_idx} according to the correponding comit certificates.

\paragraph{Remarks.} 
We remark that we did not implement the rare case when a non-TEE node becomes leader. In such a scenario, the CoSi scheme \cite{cosi} is used to aggregate authenticated messages, and the performance is expected to be worse than the case when a TEE node leads the protocol. 
In our current implementation, when a non-TEE node becomes leader, we make the protocol ``idle-waiting'' until a TEE node is available in the network.

In addition, transactions in CCF are committed in batch, rather than one by one. The maximal batch size for transcations defaults to 20,000 bytes (20 KiB). However, when the request timeout reaches, it triggers the comitting for transcations since last commit. Therefore, the batch size for transactions varies, with a maximal size of 20 KiB.

%% file: eval.tex
\section{Evaluation}
\label{sec:eval}

This section presents our experimental study of \mraft, focusing on its performance (i.e., transaction throughput and latency) and scalability.

We conducted experiments on Microsoft Azure cloud platform using SGX-enabled virtual machines (VMs) backed by Intel Xeon E-2288G processor, i.e., DCsv2-series Confidential Computing VMs \cite{azure_cc}. 
We chose size ``Standard\_DC4s\_v2'' for all VMs, each configured with 4 vCPUs, 112 MiB EPC memory, 16 GiB memory, and 30 GiB SSD.

For all experiments, we run each \mraft~node in a separate VM, running Ubuntu 18.04.5. We deploy these VMs evenly across five Azure datacenters, i.e., ``East US'', ``Canada Central'', ``UK South'', ``West Europe'', and ``Southeast Asia''.
We issue client requests to the \mraft~backed service using a ``Standard\ D8as\_v4'' VM located at the ``East US'' datacenter; this VM is configured with 8 vCPUs, 32 GiB RAM and 30 GiB SSD, running Ubuntu 20.04.2.
We report the average communication latency between these nodes in Table~\ref{tab:azure_network_latency}.

\input{tables/azure_network_latency.tex}

\paragraph{Benchmarks} We use two benchmarks in the experiments.
The first benchmark runs a \texttt{Logging} application, which supports storing a \texttt{message} with \texttt{id} and retriving the stored \texttt{message} with a given \texttt{id}. This benchmark invovles 100,000 transcations of storing messages of the form $\langle \texttt{id}, \texttt{msg} \rangle$, wherein \texttt{id} is a unique integer and \texttt{msg} is the SHA256 checksum of \texttt{id}.
Since this benchmark only involves transactions on a single table, for the second one, we use the more complicated \texttt{TPC-C} benchmark \cite{tpc-c}.
The TPC-C database is composed of nine types of tables with a wide range of row sizes and cardinalities.
TPC-C involves a mix of five concurrent transactions of different types and complexity.
Therefore, there is greater diversity in the data manipulated by the five types of transactions and thus greater database contention. In the second benchmark, we also issue 100,000 transcations.

In all experiments, transaction throughput is measured at the \textit{leader} replica and latency at the clients. 
Latency is averaged over all transactions in an experiment and counts the time from sending a command on the client to receiving a global commit conﬁrmation.
Unless otherwise stated, the results presented in this section are averaged over $10$ independent runs.
We focus on normal operation, and do not report performance of the system in case that the leader crashes or during the leader election process~\cite{raft}.

\paragraph{Baselines} We compare \mraft~against two baselines: 
(i) \texttt{PBFT}, wherein each node runs PBFT without any TEE hardening; and
(ii) \texttt{Raft-TEE}, wherein each node runs Raft within a TEE.
Note that \mraft, \texttt{PBFT} and \texttt{Raft-TEE} all aim to achieve the same goal, i.e., state machine replication in a network wherein all nodes are running within adversarial environments. A brief comparison of them is shown in Table \ref{tab:mraft_comparison}.
For all experiments, we configure the number of TEE nodes in \mraft~to be $\lfloor \frac{n}{2} \rfloor$, wherein $n$ is the network (i.e., cluster) size.
\input{tables/comparison.tex}

\input{figures/throughput.tex}
\paragraph{Experimental Results.} Now we present our evaluation results.
We remark that in all comparisons, we have normalized \rafttee's values to settings such that it has the same fault threashold $f$ with \mraft~ and \pbft.

Figure \ref{fig:mraft_throughput} presents the throughput of \mraft, \texttt{PBFT}, and \texttt{Raft-TEE} with respect to different cluster sizes ($n$) on Azure. 
As can be seen, \mraft's throughput outperform both \texttt{PBFT} and \texttt{Raft-TEE} in both benchmarks, regardless of the cluster size.

\input{figures/client_latency.tex}
Figure \ref{fig:mraft_latency} depicts the latency of \mraft, \texttt{PBFT}, and \texttt{Raft-TEE} with respect to different cluster sizes ($n$) on Azure.
Interestingly, \mraft's latency is smaller than both \texttt{PBFT} and \texttt{Raft-TEE} in both benchmarks, regardless of the cluster size.
That is, in terms of performance (i.e., transaction throughput and latency), \mraft~ outperforms both \pbft~ and \rafttee.

Next, we compare the scalability of \mraft~ with that of \pbft~and \rafttee.
As shown in Figure \ref{fig:mraft_throughput}, as the cluster size increases, the throughput of \pbft~ and \rafttee~ drops much faster than \mraft.
Similarly, as can be seen in Figure~\ref{fig:mraft_latency}, the latency of \pbft~ and \rafttee~ increases much faster than \mraft~ when the cluster size increases. 
Even when the cluster size increases to $n=47$, \mraft's throughput still does not drop that much, and its latency also does not increase that much, as compared with \pbft~ and \rafttee. 
These results demonstrated \mraft's excellent scalability.

In summary, \mraft's performance (i.e., transaction throughput and latency) outperforms both \pbft~ and \rafttee. At the same time, \mraft~ also provides better scalability than \pbft~ and \rafttee. 
That is, \textit{MFT protocols achieve the same security gurantees as their BFT counterparts, but also provide better performance and scalability}.

%% file: tables/azure_network_latency.tex
\definecolor{grad0}{RGB}{235, 130, 128} 
\definecolor{grad1}{RGB}{240, 185, 147} 
\definecolor{grad2}{RGB}{245, 226, 150} 
\definecolor{grad3}{RGB}{225, 235, 185} 
\definecolor{grad4}{RGB}{220, 245, 215}

\begin{table}
\centering

\caption{Network latency (ms) between nodes on Azure}
\label{tab:azure_network_latency}

\resizebox{0.47\textwidth}{!} {
\begin{tabular}{|l|c|c|c|c|c|}
\hline
\textbf{Datacenter} & \textbf{East US} & \textbf{Canada Central} & \textbf{UK South} & \textbf{West Europe} & \textbf{Southeast Asia} \\
\hline\hline
\textbf{East US} & \cellcolor{grad0} 1.71 & \cellcolor{grad0} 27.89  & \cellcolor{grad0} 75.34 & \cellcolor{grad0} 82.82 & \cellcolor{grad0} 219.86 \\
\hline
\textbf{Canada Central} & \cellcolor{grad1} 27.89  & \cellcolor{grad1} 3.50  & \cellcolor{grad1} 90.0  & \cellcolor{grad1} 93.94  & \cellcolor{grad1} 218.11  \\
\hline
\textbf{UK South} & \cellcolor{grad2} 75.34  & \cellcolor{grad2} 90.0  & \cellcolor{grad2} 1.27  & \cellcolor{grad2} 8.95   & \cellcolor{grad2} 156.12  \\
\hline
\textbf{West Europe} & \cellcolor{grad3} 82.82  & \cellcolor{grad3} 93.94  & \cellcolor{grad3} 8.95  & \cellcolor{grad3} 2.35 & \cellcolor{grad3} 160.39  \\
\hline
\textbf{Southeast Asia} & \cellcolor{grad4} 219.86  & \cellcolor{grad4} 218.11 & \cellcolor{grad4} 156.12  & \cellcolor{grad4} 160.39 & \cellcolor{grad4} 2.12 \\
\hline
\end{tabular}
}

\end{table}

%% file: tables/comparison.tex
\begin{table}
\caption{Comparison of \mraft~with \texttt{PBFT} and \texttt{Raft-TEE} }
\label{tab:mraft_comparison}

\centering

\resizebox{0.47\textwidth}{!} {
\begin{tabular}{|l|c|c|c|}
\hline
                 & $\texttt{MRaft}$ & $\texttt{PBFT}$ & $\texttt{Raft-TEE}$   \\ \hline
TEE Availability & Some nodes       & None       & All nodes          \\ \hline
\multicolumn{1}{|l|}{\begin{tabular}[c]{@{}c@{}}Fault-Tolerance\\ Threshold \end{tabular}}
   & $f=\frac{n-2}{3} $      & $f=\frac{n-1}{3}$        & $f=\frac{n-1}{2}$ \\ \hline
\end{tabular}
}

\end{table}

%% file: figures/throughput.tex
\definecolor{auburn}{rgb}{0.43, 0.21, 0.1}
\definecolor{britishracinggreen}{rgb}{0.0, 0.26, 0.15}

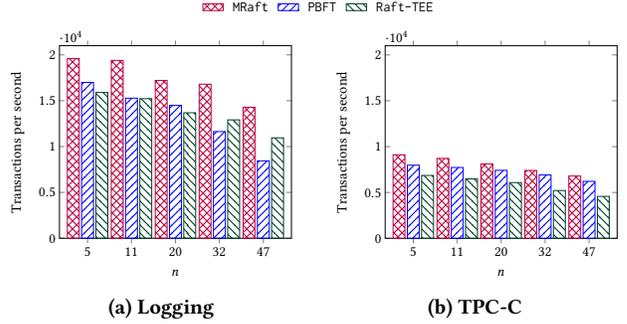
\begin{figure}
\centering 

\begin{subfigure}{.23\textwidth}
\centering
\begin{tikzpicture}[thick, scale=.45]
\begin{axis}[
	ticklabel style = {font=\LARGE},
    ybar,
    enlarge x limits = 0.16,
    legend style={at={(0.6,1.2)},anchor=west, draw=none,legend columns=3, font = \LARGE, column sep=0.2cm},
    bar width=10pt,  
    area legend,
    ylabel={Transactions per second},   
    xlabel={$n$},   
    y label style={font = \LARGE, at={(0,0.5)}},   
    x label style={font = \LARGE, at={(0.5, -0.05)}},      
    xticklabels={0,5, 11, 20, 32, 47},
    ymin = 0, ymax = 21000]

% MRaft
\addplot [color = purple,thick , pattern color = purple, pattern = crosshatch] 
coordinates { (0,19603) (1,19396) (2,17212) (3,16802) (4,14289) };

% PBFT
\addplot[color = blue,thick , pattern color = blue, pattern = north east lines] 
coordinates { (0,16990) (1,15277) (2,14488) (3,11635) (4,8442) };

% Raft-TEE
\addplot [color = britishracinggreen,thick , pattern color = britishracinggreen, pattern = north west lines] 
coordinates { (0,15901) (1,15225) (2,13685) (3,12911) (4,10949) };

\legend{ \texttt{MRaft}, \texttt{PBFT}, \texttt{Raft-TEE}}
\end{axis}
\end{tikzpicture}
\caption{Logging} 
\label{fig:mraft_tp:a}
\end{subfigure}
\begin{subfigure}{.23\textwidth}
\centering
\vspace{4.5mm}

    \begin{tikzpicture}[thick, scale=.45]
    \begin{axis}[
    ticklabel style = {font=\LARGE},
    ybar,
    enlarge x limits = 0.16,
    legend style={at={(0.65,1.3)},anchor=west, draw=none,legend columns=3, font = \LARGE, column sep=0.2cm},
    bar width=10pt,  
    area legend,
    ylabel={Transactions per second},   
    xlabel={$n$},   
    y label style={font = \LARGE, at={(0,0.5)}},   
    x label style={font = \LARGE, at={(0.5, -0.05)}},      
    xticklabels={0,5, 11, 20, 32, 47},
        ymin = 0, ymax = 21000]
    
    % MRaft
    \addplot [color = purple,thick , pattern color = purple, pattern = crosshatch] 
    coordinates { (0,9112) (1,8713) (2,8106) (3,7403) (4,6791) };
    
    % PBFT
    \addplot[color = blue,thick , pattern color = blue, pattern = north east lines] 
    coordinates { (0,7976) (1,7725) (2,7413) (3,6909) (4,6209) };

    % Raft-TEE
    \addplot [color = britishracinggreen,thick , pattern color = britishracinggreen, pattern = north west lines] 
    coordinates { (0,6854) (1,6480) (2,6077) (3,5211) (4,4573) };
    
    \end{axis}
    \end{tikzpicture}
    
\caption{TPC-C} 
\label{fig:mraft_tp:b}
\end{subfigure}

\caption{Throughput of \mraft, \texttt{PBFT} and \texttt{Raft-TEE} with respect to different cluster sizes ($n$) on Azure. Here, \rafttee~ is the case wherein each node runs Raft within a TEE.}
\label{fig:mraft_throughput}

\end{figure}

%% file: figures/client_latency.tex
\definecolor{auburn}{rgb}{0.43, 0.21, 0.1}
\definecolor{britishracinggreen}{rgb}{0.0, 0.26, 0.15}

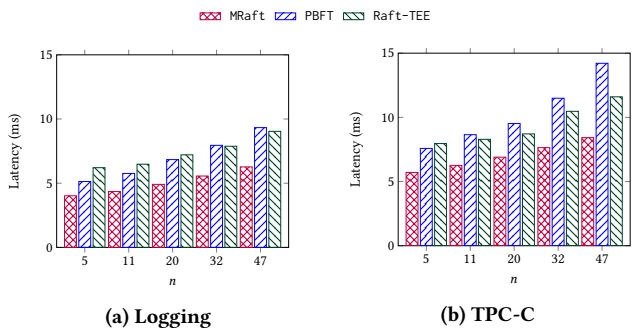
\begin{figure}
\centering 

\begin{subfigure}{.23\textwidth}
\centering
\begin{tikzpicture}[thick, scale=.45]
\begin{axis}[
	ticklabel style = {font=\LARGE},
    ybar,
    enlarge x limits = 0.16,
    legend style={at={(0.6,1.2)},anchor=west, draw=none,legend columns=3, font = \LARGE, column sep=0.2cm},
    bar width=10pt,  
    area legend,
    ylabel={Latency (ms)},   
    xlabel={$n$},   
    y label style={font = \LARGE, at={(0,0.5)}},   
    x label style={font = \LARGE, at={(0.5, -0.05)}},      
    xticklabels={0,5, 11, 20, 32, 47},
    ymin = 0, ymax = 15]

% MRaft
\addplot [color = purple,thick , pattern color = purple, pattern = crosshatch] 
coordinates { 
(0,4.02) (1,4.35) (2,4.91) (3,5.56) (4,6.27) 
};
% PBFT
\addplot[color = blue,thick , pattern color = blue, pattern = north east lines] 
coordinates { 
(0,5.13) (1,5.76) (2,6.84) (3,7.95) (4,9.33) 
};
% Raft-TEE
\addplot [color = britishracinggreen,thick , pattern color = britishracinggreen, pattern = north west lines] 
coordinates { 
(0,6.21) (1,6.48) (2,7.21) (3,7.88) (4,9.04) 
};

\legend{ \texttt{MRaft}, \texttt{PBFT}, \texttt{Raft-TEE}}
\end{axis}
\end{tikzpicture}
\caption{Logging} 
\label{fig:mraft_latency:a}
\end{subfigure}
\hfill
\begin{subfigure}{.23\textwidth}
    \centering
    \vspace{5mm}
    \begin{tikzpicture}[thick, scale=.45]
    \begin{axis}[
      ticklabel style = {font=\LARGE},
    ybar,
    enlarge x limits = 0.16,
    bar width=10pt,  
    area legend,
    ylabel={Latency (ms)},   
    xlabel={$n$},   
    y label style={font = \LARGE, at={(0,0.5)}},   
    x label style={font = \LARGE, at={(0.5, -0.05)}},      
    xticklabels={0,5, 11, 20, 32, 47},
        ymin = 0, ymax = 15]

    % MRaft
    \addplot [color = purple,thick , pattern color = purple, pattern = crosshatch] 
    coordinates { 
    (0,5.71) (1,6.26) (2,6.9) (3,7.64) (4,8.42) 
    };
    
    % PBFT
    \addplot[color = blue,thick , pattern color = blue, pattern = north east lines] 
    coordinates { 
    (0,7.58) (1,8.65) (2,9.52) (3,11.49) (4,14.21) 
    };

    % Raft-TEE
    \addplot [color = britishracinggreen,thick , pattern color = britishracinggreen, pattern = north west lines] 
    coordinates { 
    (0,7.96) (1,8.29) (2,8.71) (3,10.47) (4,11.60) 
    };

    \end{axis}
    \end{tikzpicture}
    \caption{TPC-C} 
    \label{fig:mraft_latency:b}
    \end{subfigure}

\caption{Latency of \mraft, \texttt{PBFT} and \texttt{Raft-TEE} with respect to different cluster sizes ($n$) on Azure. Here, \rafttee~ is the case wherein each node runs Raft within a TEE.}
\label{fig:mraft_latency}
\end{figure}

%% file: related_work.tex
\section{Related Work}
\label{sec:related_work}

The bottleneck of performance (i.e., transaction throughput and latency) and scalability in blockchain systems or distributed ledger systems is typically the underlying consensus protocol.
Consensus protocols are used by replicas to agree on an order for transactions. 
A majority of current ledger systems \cite{hyperledger_fabric,goquorum} rely on BFT consensus protocols.

\paragraph{Improving BFT Protocols.} Several recent works have improved the scalability of BFT protocols. Using threshold cryptograph, SBFT \cite{sbft} proposes a variant of PBFT that scales to larger consensus groups. Byzcoin \cite{byzcoin} also builds on PBFT and dynamically forms consensus groups. HotStuff \cite{hotstuff} can also scale to hundreds of replicas using threshold cryptography.

\paragraph{Improving Consensus Protocols using TEE} Several works have proposed to improve the efficiency of BFT protocols using TEE \cite{hypster,a2m,ccf_tr}. 
These systems typically assume that each node is equipped with a small trusted subsystem that fails only by crashing, whereas other untrusted components in a node may fail or misbehave arbitarily. 
The use of such trusted subsystems reduces the number of requried nodes to tolerate $f$ failures. 
However, this line of protocols impose a trust assumption on each and every node participating in the system, which may not be applicable to settings wherein participants have different hardware conﬁgurations. 
Unlike existing works, we explore a new approach to designing efﬁcient distributed fault-tolerant systems that tolerate a combination of crash and Byzantine faults, which we refer to as mixed fault tolerance (MFT).

%% file: conclusion.tex
\section{Conclusion}
\label{sec:conclusion}

We proposed a new approach, which leverages TEE, to designing efficient distributed fault-tolerant protocols (i.e., MFT protocols) that tolerate a combination of crash and Byzantine faults.
We identified two key principles for designing efficient MFT protocols, and showcased these two principles by prescribing an MFT protocol, namely, \mraft. 
We implemented a prototype of \mraft, integrated it into the CCF \cite{ccf_tr} blockchain framework, conducted experiments in realistic deployment settting, and demonstrated the efficiency of our approach.